# Mains-Synchronized Timing Trigger for Stability Enhancement in FEL Pulsed Microwave Systems

Jinfu Zhu[a]*, Hongli Ding[b]*, Haokui Li[b], Qiaoye Ran[a], Jiayue Yang[b], Weiqing Zhang[b]

a: Institute of Advanced Science Facilities, Shenzhen, China

b: Dalian Institute of Chemical Physics, Chinese Academy of Sciences, Dalian, China

## Abstract

Pulsed microwave stability in linear accelerators (LINACs) is critical for maintaining high-quality electron beams in synchrotron radiation and free-electron laser (FEL) facilities. This study establishes and validates a zero-crossing synchronization strategy to suppress mains-induced disturbances in the Dalian Coherent Light Source (DCLS) accelerator. Through comprehensive numerical simulations at 10 Hz, 20 Hz, and 25 Hz repetition rates, we first demonstrate the temporal evolution of microwave amplitude perturbations under mains power disturbances. Experimental validation with the digital Low-Level Radio Frequency (LLRF) system reveals a substantial similarity between simulated and measured interference patterns, confirming the susceptibility of the klystron output to mains power modulation. The developed synchronization technique integrates real-time mains zero-crossing detection with precision timing sequence generation, effectively decoupling the microwave system from power frequency fluctuations. It achieves remarkable inter-pulse stability improvements, reducing microwave amplitude fluctuation from ~0.30% RMS to 0.07% RMS. This approach not only addresses the inherent bandwidth limitations of conventional proportional-integral-derivative (PID) controllers in low-repetition-rate accelerators but also provides a solution for troubleshooting power-related instabilities in advanced light source facilities.

## 1. Introduction

Synchrotron radiation facilities and free-electron lasers (FELs) have revolutionized modern scientific research by enabling atomic-scale spatial resolution and femtosecond temporal resolution for probing dynamic processes [1-3]. The evolution of light source technology has achieved a ten-order-of-magnitude improvement in spectral brilliance compared to first-generation synchrotrons [4], with modern FELs facilitating unprecedented observations of molecular dynamics [5-7]. These advancements critically depend on linear accelerators (LINACs) capable of generating electron beams with sub-micron emittance and ultra-low energy spread, for which microwave inter-pulse stability in pulsed radiofrequency (RF) systems becomes essential. Inter-pulse stability refers to the consistency of microwave parameters (e.g., amplitude and phase) between consecutive pulses, a property critical for maintaining beam quality in pulsed accelerators. Experimental studies at REGAE and SwissFEL demonstrate that sub-0.02% amplitude stability is essential to maintain electron beam energy uniformity and peak current control, directly enabling femtosecond-scale temporal resolution and high-brilliance FEL output [8,9]. For facilities like PAL-XFEL and the European XFEL, even ppm-level deviations in RF amplitude critically degrade performance: PAL-XFEL requires <0.02% amplitude stability to ensure 10 GeV beam energy consistency [10], while the European XFEL achieves 0.05 nm wavelength operation by maintaining 0.01% amplitude stability [11]. Therefore, precise RF control is indispensable for preserving beam stability and enabling many FEL applications.

However, the implementation challenges differ significantly between accelerator types. While continuous-wave (CW) accelerators can correct perturbations in real-time through closed-loop feedback [12], pulsed accelerators face distinct limitations due to their low-duty-cycle characteristics (<0.1%). Specifically, the inherently restricted closed-loop bandwidth of conventional proportional-integral-derivative (PID) controllers in pulsed systems [13] leads to inadequate suppression of high-frequency disturbances. This limitation stems from a fundamental trade-off between sampling rate and controller bandwidth: the PID's effective noise suppression range is typically confined to frequencies below ~1/5 of the closed-loop bandwidth [14]. For instance, a system with a 1 kHz closed-loop bandwidth can only suppress disturbances below ~200 Hz. In low-duty-cycle accelerators operating at 10-50 Hz repetition rates, the controller update interval (20-100 ms) further constrains the achievable closed-loop bandwidth to sub-100 Hz levels, leaving high-frequency mains-induced noise components (e.g., 50/60 Hz harmonics or aliased signals) beyond the controller's attenuation capability. Consequently, residual modulation of inter-pulse microwave parameters persists, a phenomenon exacerbated when mains power frequency (50/60 Hz) noise couples into klystron modulators through AC/DC converters. Such coupling generates beat frequency effects with microwave pulse repetition rates, causing periodic modulation of inter-pulse microwave parameters.

*Corresponding author, e-mail: zhujinfu@mail.iasf.ac.cn, dinghongli@dicp.ac.cn.

These challenges are exemplified in the Dalian Coherent Light Source (DCLS), China's first extreme ultraviolet FEL facility based on the high-gain harmonic generation (HGHG) principle. The DCLS achieves wavelength-tunable operation (50–150 nm) at up to 50 Hz repetition rates, a capability critical for molecular photoionization dynamics studies [15,16]. However, during its Phase-II upgrades, significant microwave modulation phenomena were observed. Notably, a ~9.95 Hz perturbation emerged during 20 Hz operation, inducing microwave amplitude jitter exceeding 0.3% RMS and degrading electron beam energy stability beyond design specifications (0.10%→0.30%). Although recent advances in active noise control (ANC) [17] and feedforward compensation algorithms [18] offer partial solutions, their effectiveness against such rapid perturbations remains limited. Recent studies on FEL timing systems emphasize the critical need for robust synchronization architectures to mitigate phase uncertainties caused by power cycles and frequency dividers [19, 20]. For instance, the trigger signal for master oscillators in FEL systems is typically derived from AC mains zero-crossing detection [21], synchronizing the entire machine to grid fluctuations and minimizing inter-pulse deviations.

This paper establishes a zero-crossing synchronization strategy that phase-locks microwave pulse acquisition to mains power zero-crossing phases using timing signals from synchronization systems, thereby addressing both the mechanistic origins of aliased disturbances and their practical mitigation in DCLS. Through combined simulations of mains interference mechanisms and experimental validation at DCLS, we systematically reveal the modulation mechanisms of mains frequency disturbances on microwave parameters and demonstrate the robustness of our approach across 10–25 Hz repetition rates. Implementation at DCLS reduced microwave amplitude fluctuations from ~0.30% RMS to 0.07% RMS. This work integrates numerical modeling of interference mechanisms with multi-rate experimental validation, thereby establishing a universal framework for power-line noise suppression in advanced light sources.

## 2. Methodology

### 2.1 System Architecture

As illustrated in Fig. 1, the DCLS system architecture consists of two primary physical systems: a linear electron accelerator and an FEL amplifier system, along with auxiliary subsystems. The entire facility spans 100 meters within a radiation-shielded tunnel and integrates advanced laser systems (e.g., photocathode RF-Gun and seed laser), accelerator components (e.g., S-band modules), and experimental infrastructure. The driver laser and seed laser operate synergistically with the electron beam dynamics. The driver laser, operating within the photocathode RF-Gun, generates high-quality electron beams via photoemission, ensuring low emittance and precise temporal control. The seed laser, with tunable wavelengths (240–360 nm) and a peak power of up to 100 MW, initiates the HGHG process in the modulator undulator, imprinting energy modulation onto the electron beam. The experimental station utilizes amplified EUV pulses (50–150 nm, >100 MW peak power, 30 fs–1 ps duration) for ultrafast studies in molecular dynamics, surface science, and quantum materials. These experiments are supported by diagnostics and beamline optics tailored for high-resolution measurements.

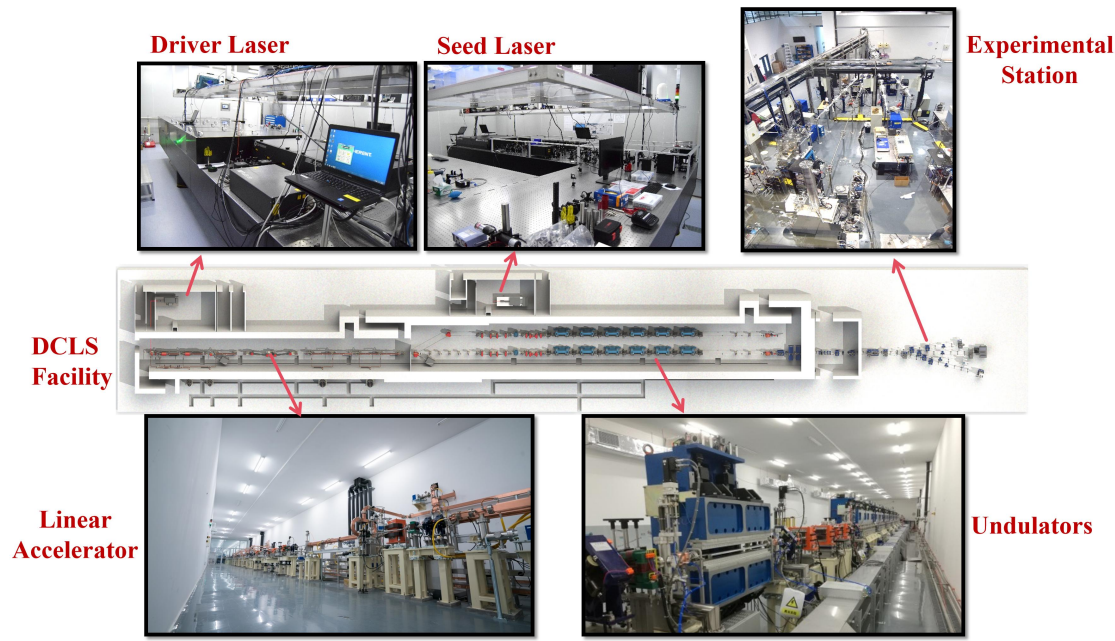

Figure 1. The system architecture of the DCLS facility

The DCLS LINAC and its microwave system, depicted in Fig. 2, are designed to ensure stable electron beam acceleration and synchronization. A master oscillator (2856 MHz) generates the fundamental microwave signal, which is distributed via a reference distribution system to maintain phase coherence across the S-band accelerating modules (A0–A6). Four independent microwave systems are designed to provide stable microwave excitation for the electron gun and modules A0–A6. Each system comprises a low-level radiofrequency (LLRF) subsystem, a solid-state amplifier (SSA), a klystron with high-voltage modulation, and waveguide components. The LLRF system generates phase-locked microwave signals synchronized with the master oscillator. These signals undergo two-stage amplification via the SSA and klystron to deliver high-power microwaves at 50–80 MW. The LLRF system implements real-time feedback or feedforward control by monitoring the klystron output, ensuring field stability and phase accuracy.

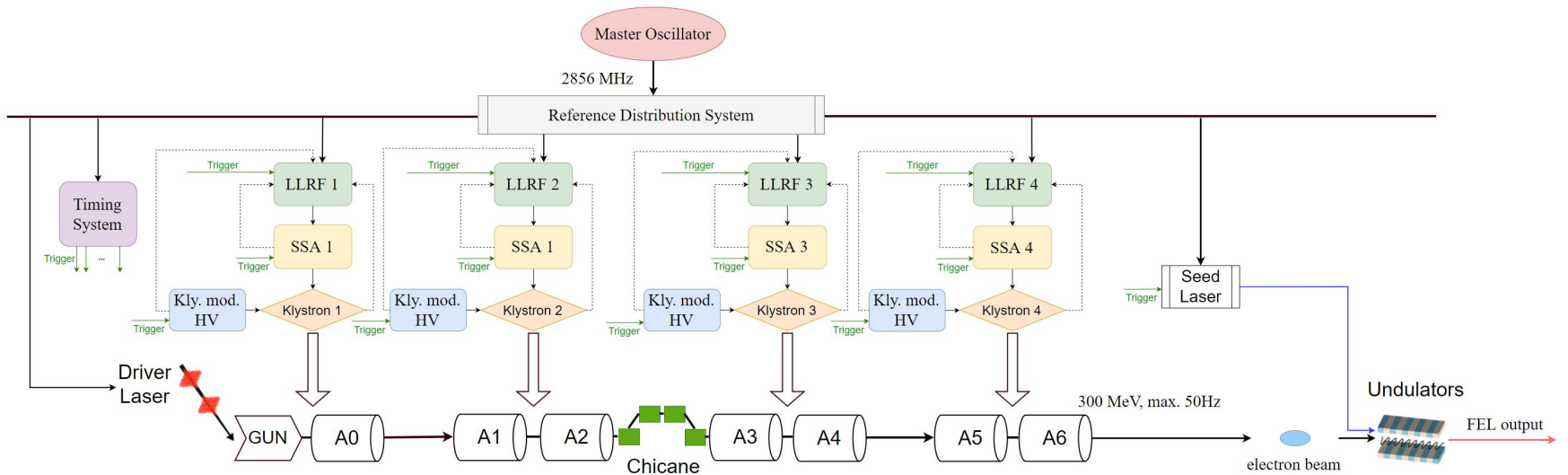

Figure 2. The DCLS LINAC and its microwave system

The accelerator is divided into two functional segments: a high-brightness photocathode injector (RF-Gun and A0) for low-emittance beam generation, and a main accelerating section (A1–A6) to boost energy and compress the beam longitudinally. The magnetic compressor (BC), located at the end of A2, employs four dipole magnets to shape the electron beam into an under-compressed state with a head-to-tail energy chirp. Subsequent accelerating modules (A3–A6) operate at off-crest phases to mitigate energy spread through a combination of radiofrequency (RF) acceleration and longitudinal wakefield effects. At the LINAC exit, the electron beam reaches an energy of ~300 MeV with a maximum repetition rate of 50 Hz. This beam is synchronized with the seed laser in the undulator via a precision timing network, achieving sub-picosecond coordination. The interaction between the microbunched electron beam and the seed laser generates intense, coherent FEL radiation. This capability enables applications in ultrafast spectroscopy and high-resolution imaging.

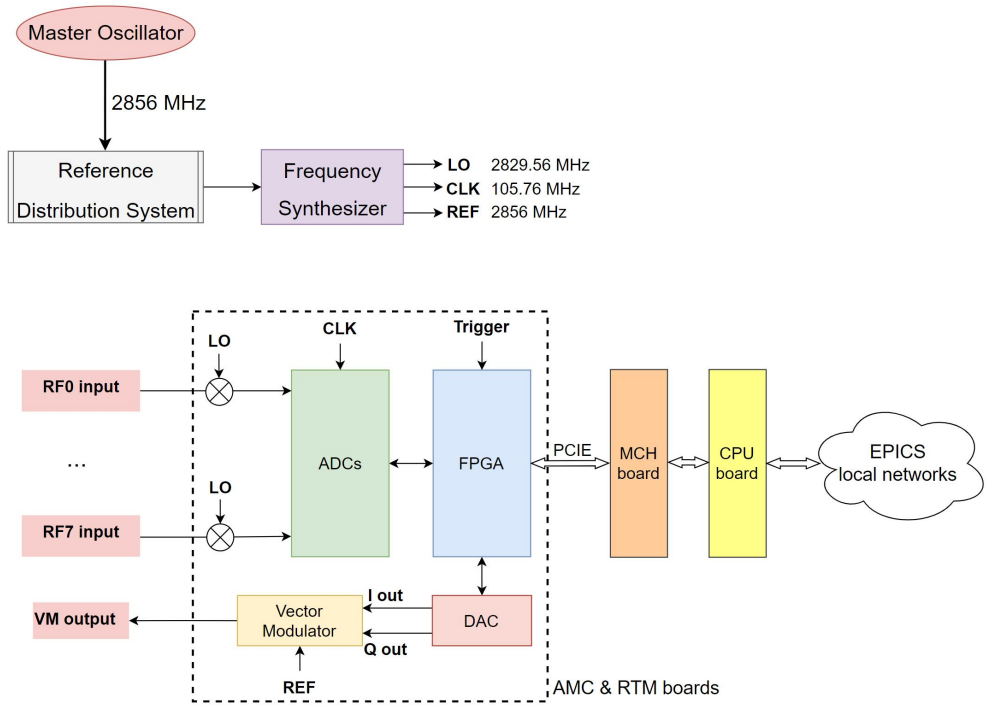

Figure 3. Timestamp receiver IP core firmware composition diagram

As shown in Fig. 3, the LLRF system acts as the control core of the microwave system. It processes the 2856 MHz master oscillator signal through a frequency synthesizer to generate three critical clocks: a local oscillator (LO) at 2829.56 MHz, a sampling clock (CLK) at 105.76 MHz for analog-to-digital converters (ADCs), and a reference clock (REF) at 2856 MHz. External radiofrequency (RF) input signals (RF0–RF7) are down-converted to an intermediate frequency (IF) of 26.44 MHz through mixing with the LO signal. The IF signals are digitized by ADCs into vectors $I$ and $Q$. A quadruple sampling technique is employed to sequentially generate vectors $I, -Q, -I, Q$, as defined by Eq. (2.1). These vectors are subsequently used to calculate the amplitude $A$ through Eq. (2.2). The single-shot microwave pulse width of the DCLS klystron is approximately 2.5 μs. Considering the filling time requirement of approximately 800 ns for the traveling-wave tube system, the RF amplitude characterization is performed through time-domain integration over an 800 ns window (e.g. 1.6–2.4 μs) to ensure stable field establishment. Since the microwave pulse width (2.5 μs) is significantly shorter than the mains cycle period (20 ms for 50 Hz), temporal integration of sampled voltages inherently superimposes mains signal components, leading to systematic measurement errors. Figure 4(a) shows the typical directly digitized IF signal, while Fig. 4(b) presents the corresponding quadrature vectors $I/Q$ and the RF amplitude waveform.

$$A\cos(\omega k T_s + \varphi) = I\cos\left(\frac{k\pi}{2}\right) - Q\sin\left(\frac{k\pi}{2}\right) = I, -Q, -I, Q, \ldots \text{ where } T_s = \frac{1}{4} * \frac{2\pi}{\omega}, k = 0,1,2\ldots \quad (2.1)$$

$$A = \sqrt{I^2 + Q^2} \quad (2.2)$$

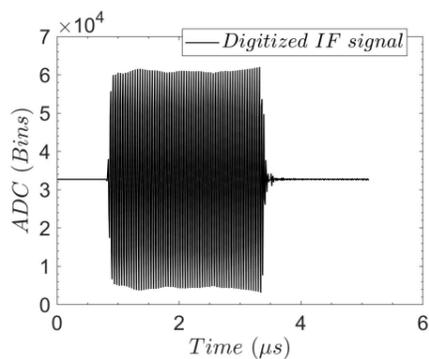

(a) Time-domain raw digitized IF signal with high-frequency oscillations

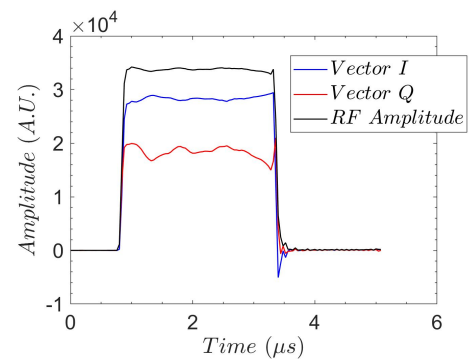

(b) Quadrature vectors (I/Q) and amplitude waveform after down-conversion

Figure 4. Microwave signal processing of the DCLS klystron

User-defined amplitude and phase settings are converted via Eq. (2.3) into I/Q modulation signals. These signals are synthesized by digital-to-analog converters (DACs) and upconverted to microwave excitation signals through a vector modulator (VM).

$$I = A_{user}\cos(\varphi_{user}), Q = -A_{user}\sin(\varphi_{user}) \quad (2.3)$$

All signal processing algorithms-including feedback control and I/Q modulation-run on a field-programmable gate array (FPGA) to enable real-time operation. The DCLS LLRF system is based on the Micro Telecommunications Computing Architecture (MTCA.4) platform. Rear transition module (RTM) boards perform analog signal conditioning, while advanced mezzanine card (AMC) boards manage waveform digitization. The MicroTCA Carrier Hub (MCH) board handles chassis-level communication and data routing, and the CPU board executes embedded control software on a Linux-based operating system. The Experimental Physics and Industrial Control System (EPICS) framework is hosted on the CPU to enable remote monitoring and parameter adjustment via networked interfaces.

## 2.2. Mains-Induced Disturbance Analysis

The disturbance induced by mains frequency on microwave signals can be mathematically modeled by Eq. (2.4), where $A\cos(2\pi f t + \varphi)$ represents the original microwave signal with amplitude $A$, frequency $f$, and phase $\varphi$. The mains disturbance, characterized by amplitude $A_p$, frequency $f_p$, and phase $\varphi_p$, modulates the amplitude of the microwave signal, resulting in the perturbed output $y^*$ expressed as:

$$y^* = A_p\cos(2\pi f_p t + \varphi_p) \cdot A\cos(2\pi f t + \varphi) \quad (2.4)$$

Globally, mains frequencies are standardized at either 50 Hz (e.g., China and Europe) or 60 Hz (e.g., Japan and North America). In China, the standard mains frequency is 50 Hz with an allowable deviation of ±0.2 Hz. Figure 5 shows the time-domain simulation of the microwave signal subjected to mains-induced amplitude modulation. Figure 5(a) presents the simulated time-domain waveform of the klystron-generated microwave signal (normalized amplitude) at 2856 MHz, where high-frequency oscillations are sampled at nanosecond resolution (time in ns). Figure 5(b) displays the simulated 50 Hz mains-induced disturbance signal (normalized amplitude), plotted on a millisecond timescale (time in ms). The interaction between these signals is demonstrated in Fig. 5(c), where the modulated microwave signal (normalized amplitude, $y^*$) exhibits a distinct 50 Hz amplitude envelope that aligns with the mains frequency waveform in Fig. 5(b). This modulation originates from the coupling of mains interference into the microwave system, imprinting its periodic characteristics onto the microwave amplitude. Consequently, the LLRF system captures the modulated microwave signals, inherently reflecting the mains interference.

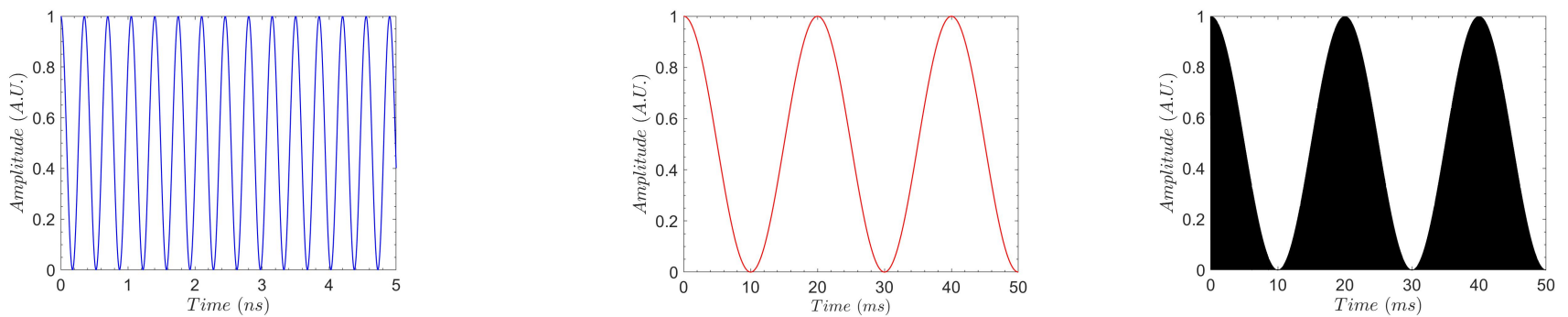

(a) Klystron-generated microwave signal (2856 MHz, normalized)  (b) 50 Hz mains disturbance signal (normalized) (c) Modulated microwave signal with 50 Hz envelope

Figure 5. Time-domain simulation of mains-induced amplitude modulation

For accelerators operating at repetition frequency $f_s$, the discretized disturbance captured by the LLRF system is governed by Eq. (2.5):

$$y_s^*[k] = A_p\cos\left(2\pi k\frac{f_p}{f_s} + \varphi_s^*\right), k = 0,1,2,\cdots \quad (2.5)$$

Here, $y_s^*[k]$ represents the per-pulse microwave amplitude perturbation, $k$ is the discrete sampling index, and $f_p/f_s$ governs the aliasing effect of mains disturbances. The perturbation frequency $f_s^*$, induced by mains interference, is derived from the Nyquist-Shannon sampling theorem and expressed as Eq. (2.6).

$$f_s^* = |f_p - N \cdot f_s|, \text{where } f_s^* \leq 0.5 f_s \; (N = 0,1,2,\ldots) \quad (2.6)$$

The spectrum of $y_s^*[k]$ contains aliased frequency components $f_s^*$ (e.g., mirrored frequencies due to undersampling), which originate from the interaction between the mains frequency $f_p$ and the accelerator repetition rate $f_s$. The aliased components $f_s^*$ dominate the perturbation patterns observed in both simulations and experiments (see Section 3.2).

## 2.3 Mains Synchronization Method

Based on the analysis of mains-induced amplitude modulation (Section 2.2), the AC zero-crossing instants are identified as optimal synchronization points since the mains-induced perturbations on microwave amplitude reach their minimum values at these temporal positions. The timing system of the DCLS employs an event-driven architecture based on an Event Generator/Event Receiver (EVG/EVR) framework. As shown in Fig. 6, the system synchronizes the master oscillator's 2856 MHz reference clock, which is divided by a frequency synthesizer to 79.33 MHz. The EVG generates timing pulse sequences with a jitter of less than 200 ps, ensuring precise synchronization across subsystems. These sequences are interlocked with the Machine Protection System (MPS) and Personal Protection System (PPS) to guarantee operational safety. The EVG outputs are distributed via fan-out modules to multiple EVRs, enabling coordinated execution of subsystems such as the microwave system (including the LLRF system, SSA, and klystron modulator), laser system, and beam diagnostics. This hierarchical timing distribution ensures compliance with the stringent temporal requirements for FEL generation.

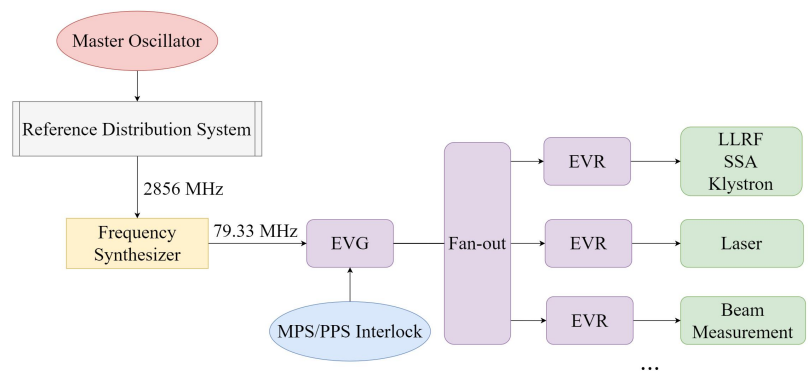

Figure 6. Schematic of the EVG/EVR-based timing distribution system.

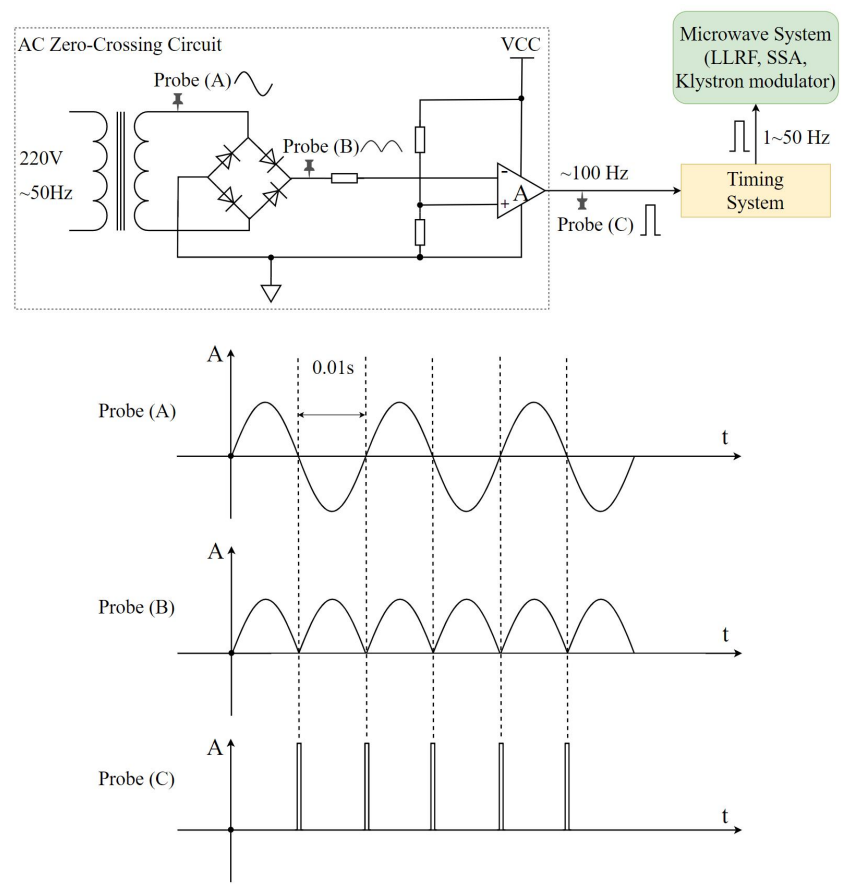

Figure 7. AC zero-crossing detection circuit and synchronized trigger generation.

The proposed mains synchronization method integrates AC zero-crossing detection with the accelerator's existing timing system. As illustrated in Fig. 7, this approach triggers the pulsed microwave system (LLRF, SSA, and klystron modulator) precisely at the zero-crossing instants of the 50 Hz mains power supply while aligning with the accelerator's repetition rate (1–50 Hz). The zero-crossing detection circuit comprises three functional stages: Probe (A) isolates and steps down the 220 V AC supply via a transformer; Probe (B) rectifies the signal to a unipolar waveform using a full-wave bridge rectifier; and Probe (C) generates a 100 Hz pulse train through a voltage comparator, synchronized to the zero-crossing points. The 100 Hz pulse is fed into the timing system, where a logical AND operation filters coherent triggers aligned with the reference signal. Subsequent frequency division generates repetition rates from 1 Hz to 50 Hz, ensuring phase-locked operation of the microwave system. This synchronization minimizes disturbances caused by power grid fluctuations, effectively decoupling the microwave system from mains-induced amplitude modulation.

## 3. Experimental Implementation and Results

### 3.1 Experimental Setup

To validate the theoretical analysis of aliased perturbation frequencies (Section 2.2), the experimental validation was performed on the DCLS LINAC at repetition rates of 10 Hz, 20 Hz, and 25 Hz. A high-precision handheld frequency meter recorded the real-time mains frequency ($f_p = 50.05 \pm 0.02$ Hz), while the digital LLRF system acquired per-pulse microwave amplitudes from Klystron 4. Data acquisition was standardized with identical sample sizes (6,000 pulses per test) to ensure statistical consistency.

Using Eq. (2.5), the measured $f_p$ was input into simulations to predict mains-induced modulation patterns. Theoretical analysis via Eq. (2.6) revealed distinct aliased perturbation frequencies:

At $f_s = 10$ Hz ($N = 5$): $f_s^* = |50.05 - 5 \cdot 10| = 0.05$ Hz

At $f_s = 20$ Hz ($N = 3$): $f_s^* = |50.05 - 3 \cdot 20| = 9.95$ Hz

At $f_s = 25$ Hz ($N = 2$): $f_s^* = |50.05 - 2 \cdot 25| = 0.05$ Hz

### 3.2 Simulation and Experimental Validation

Simulated time-domain microwave amplitudes at 10 Hz, 20 Hz, and 25 Hz (Fig. 8a–c) exhibited distinct modulation patterns: single-sinusoidal waveforms with a 20 s period dominated at 10 Hz and 25 Hz, while a dual-sinusoidal waveform emerged at 20 Hz. Corresponding frequency spectra (Fig. 8d–f) confirmed aliased components at $f_s^* = 0.05$ Hz (10 Hz/25 Hz) and 9.95 Hz (20 Hz), aligning with theoretical predictions from Eq. (2.6).

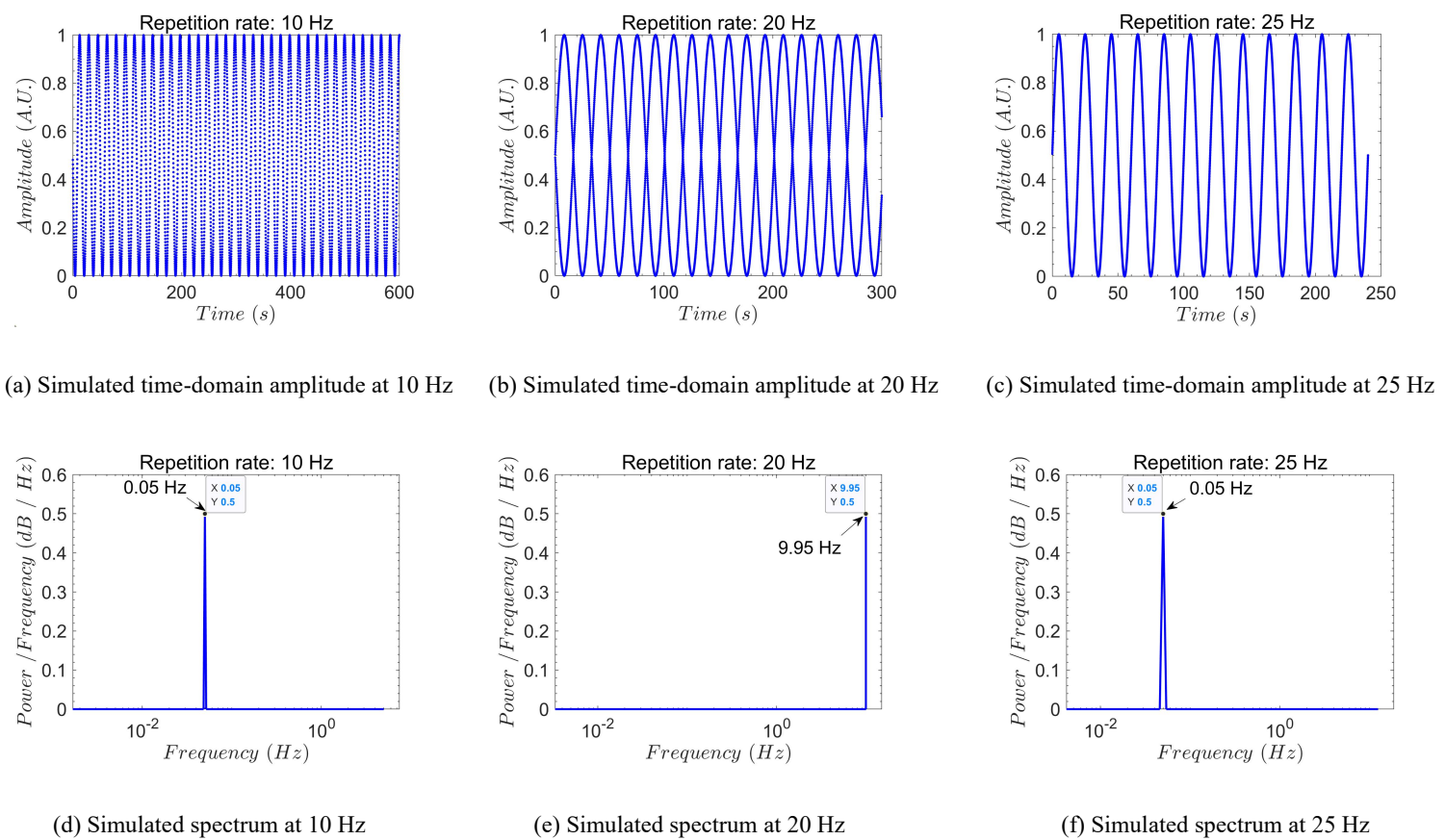

(a) Simulated time-domain amplitude at 10 Hz  (b) Simulated time-domain amplitude at 20 Hz  (c) Simulated time-domain amplitude at 25 Hz

(d) Simulated spectrum at 10 Hz  (e) Simulated spectrum at 20 Hz  (f) Simulated spectrum at 25 Hz

Figure 8. Simulated amplitude perturbations and spectra under mains interference.

Experimental measurements of microwave amplitudes (Fig. 9a–c) closely matched simulations. At 10 Hz (Fig. 9a), the time-domain waveform displayed a single-sinusoidal modulation (period $T_s^* = 1/f_s^* \approx 20$ s), while its spectrum (Fig. 9d) exhibited a peak at $0.05 \pm 0.02$ Hz, consistent with the simulated $f_s^* = 0.05$ Hz. Similarly, at 25 Hz (Fig. 9c), the modulation period and spectral peak ($0.05 \pm 0.02$ Hz) agreed with predictions. For 20 Hz (Fig. 9b), the dual-sinusoidal waveform and its spectrum (Fig. 9e) showed a dominant peak at $9.95 \pm 0.02$ Hz, validating spectral aliasing effects. Minor discrepancies in spectral peak locations ($\pm 0.02$ Hz) arose from inherent fluctuations in the mains frequency. Slow drifts in amplitude waveforms (Fig. 9a–c) were attributed to temperature and humidity variations in RF components.

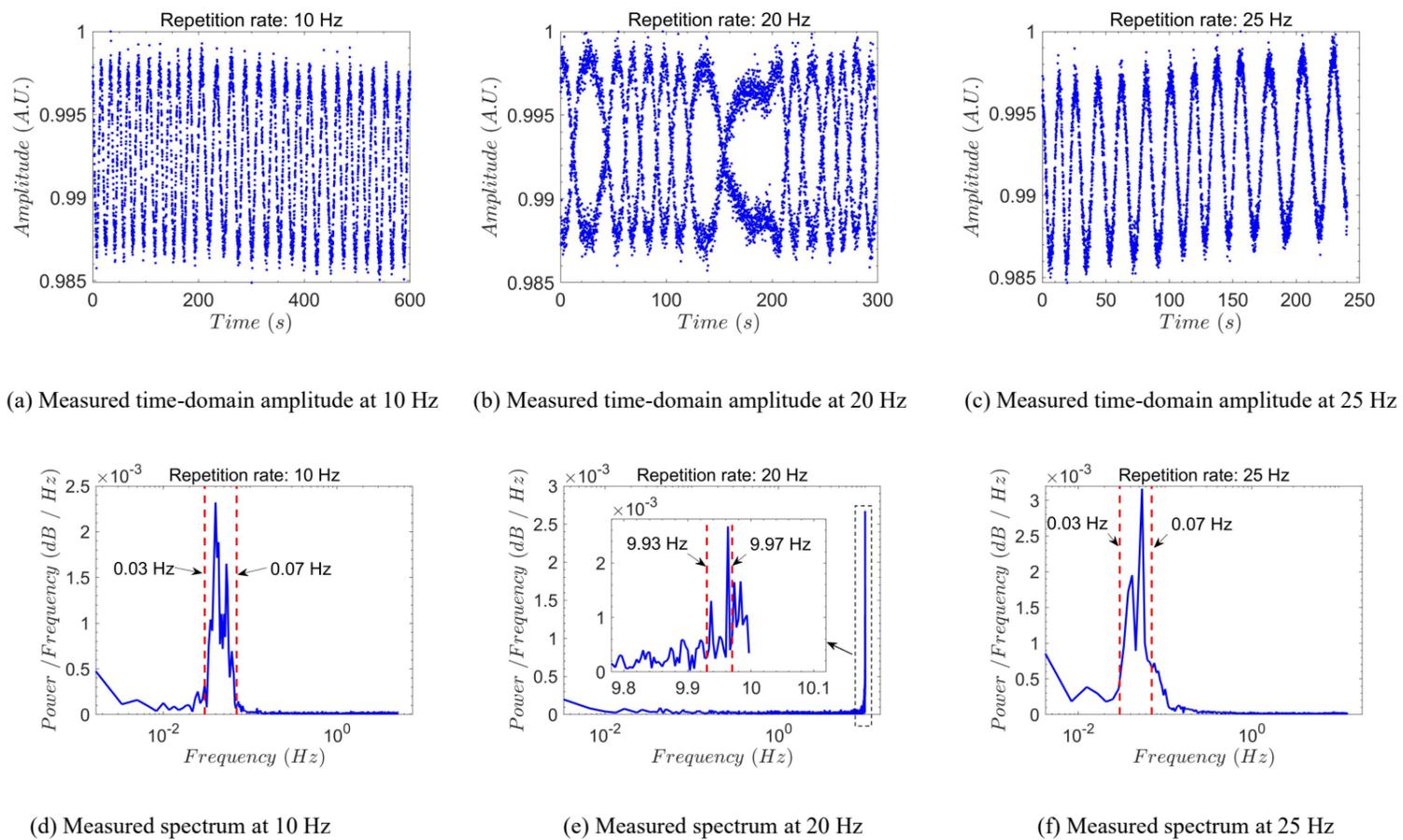

Figure 9. Experimental amplitude modulations and spectra without synchronization.

### 3.3 Performance of Synchronization Method

The proposed mains-synchronized timing trigger system integrated real-time zero-crossing detection with precision timing sequence generation. As shown in Fig. 10a–c, sinusoidal modulations were eliminated across all repetition rates. For 10 Hz and 25 Hz, the amplitude fluctuation RMS decreased from $(0.34 \pm 0.05)\%$ (unsynchronized) to $(0.07 \pm 0.01)\%$ (synchronized). At 20 Hz, dual-sinusoidal noise was suppressed from $(0.32 \pm 0.04)\%$ to $(0.07 \pm 0.01)\%$, demonstrating robustness against aliased disturbances. The synchronization method achieves phase-locked operation, ensuring microwave pulses are consistently triggered at mains zero-crossing points. The stability improvements were sustained over 24-hour tests, validating the technique's applicability in large-scale facilities.

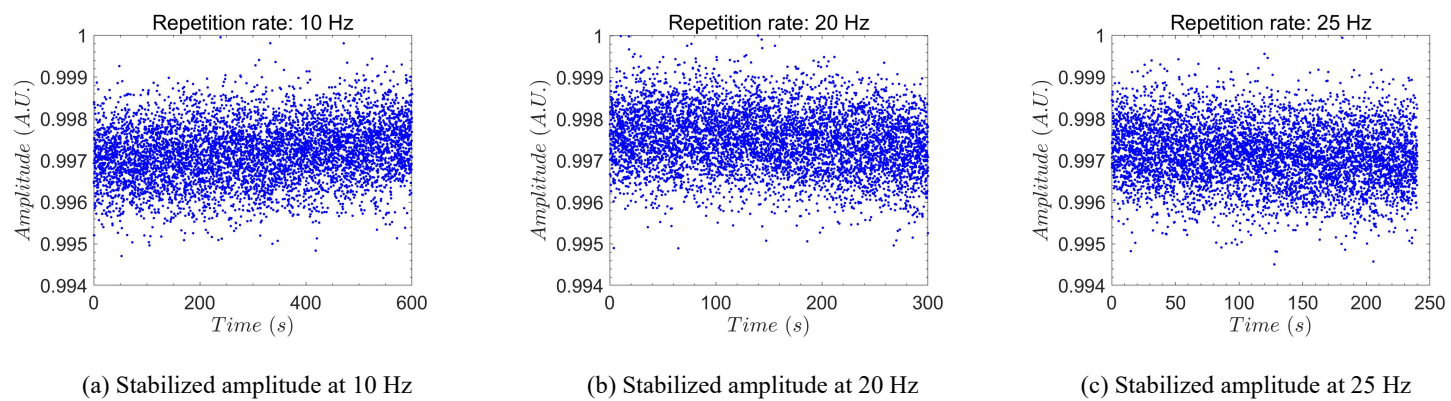

Figure 10. Amplitude stabilization using synchronization method: (a–c) Time-domain amplitudes at 10 Hz, 20 Hz, and 25 Hz, showing RMS reduction from 0.30% to 0.07%.

## 4. Conclusion and Future work

This study establishes a robust mains-synchronized timing trigger strategy to mitigate power-line-induced instabilities in pulsed microwave systems for FEL facilities. By integrating real-time zero-crossing detection of the mains power supply with precision timing sequence generation, the proposed method effectively decouples microwave amplitude fluctuations from 50 Hz grid disturbances. Numerical simulations and experimental validations at 10–25 Hz repetition rates demonstrate that the synchronization technique suppresses aliased modulation effects caused by undersampling of mains interference. At the DCLS, implementation of this approach reduced microwave amplitude jitter from 0.30% RMS to 0.07% RMS, achieving sub-0.1% inter-pulse stability critical for maintaining electron beam energy uniformity and high-brilliance FEL output. The method overcomes the inherent bandwidth limitations of conventional PID controllers in low-repetition-rate accelerators, offering a universal framework for mitigating power-line noise in advanced light sources.

While the current study focuses on repetition rates up to 25 Hz, future work could extend this technique to higher frequencies (e.g., 50 Hz) or explore its integration with other algorithms to further enhance noise suppression. Additionally, a critical direction for future research involves theoretical and experimental investigations into the impact of mains frequency noise on phase jitter. Although this work addresses amplitude modulation mechanisms comprehensively, the coupling between power-line disturbances and microwave phase stability remains unaddressed. A systematic analysis of phase jitter dynamics, supported by time-domain simulations and spectral characterization, would provide deeper insights into dual-parameter (amplitude and phase) noise coupling mechanisms. Such studies could inform the development of hybrid control strategies to simultaneously suppress both amplitude and phase fluctuations, further advancing the precision of RF systems. Moreover, the methodology holds promise for broader applications in particle accelerators, pulsed power systems, and industrial facilities requiring stringent synchronization with grid power.

## Acknowledgement

This work is supported by the National Natural Science Foundation of China (Grant No. 12405221), and the Shenzhen Science and Technology Program (Grant No. RCBS20221008093247072). We would like to thank those who collaborated on the DCLS and S$^3$FEL.

## References

[1] H. Winick, Synchrotron Radiation Sources: A Primer, World Scientific, Singapore, 1995.

[2] B.W.J. McNeil, N.R. Thompson, Nat. Photonics 4 (2010) 814-821. https://doi.org/10.1038/nphoton.2010.239

[3] F. Krausz, M. Ivanov, Rev. Mod. Phys. 81 (2009) 163-234. https://doi.org/10.1103/RevModPhys.81.163

[4] P. Emma et al., Nat. Photonics 4 (2010) 641-647. https://doi.org/10.1038/nphoton.2010.176

[5] C. Bostedt et al., Rev. Mod. Phys. 88 (2016) 015007. https://doi.org/10.1103/RevModPhys.88.015007

[6] Y. Xie et al., Science 368(6492), 767-771 (2020). https://doi.org/10.1126/science.abb1564

[7] H.S. Kang et al., Nat. Photonics 11 (2017) 708-713. https://doi.org/10.1038/s41566-017-0029-8

[8] M. Hoffmann et al., Precision LLRF Controls for the S-Band Accelerator REGAE, in Proc. IPAC'13, Shanghai, China, 2013, paper THPEA031.

[9] Z. Geng, Beam-based optimization of SwissFEL low-level RF system, Nucl. Instrum. Methods Phys. Res. A 29 (2018) 128.

[10] J. Hu et al., LLRF development for PAL-XFEL, in Proc. Int. Part. Accel. Conf. (IPAC'16), Busan, Korea, 2016, pp. 2761–2764.

[11] J. Branlard et al., The European XFEL LLRF system, in: Proc. IPAC'12, New Orleans, LA, USA, 2012, pp. 55–57.

[12] J. Branlard et al., Nucl. Instrum. Methods Phys. Res. A 732 (2013) 34-38. https://doi.org/10.1016/j.nima.2013.05.097

[13] D.E. Rivera et al., Ind. Eng. Chem. Process Des. Dev. 25 (1986) 252-265. https://doi.org/10.1021/i200032a041

[14] K.J. Åström, T. Hägglund, PID Controllers: Theory, Design, and Tuning, 2nd ed., ISA, 1995.

[15] Y. Yu et al., Chin. J. Lasers 46 (2019) 0100005. https://doi.org/10.3788/CJL201946.0100005

[16] D. Yuan et al., Science 362 (2018) 1289-1293. https://doi.org/10.1126/science.aav1356

[17] F. Qiu et al., Nucl. Instrum. Methods Phys. Res. A 955 (2020) 163280. https://doi.org/10.1016/j.nima.2019.163280

[18] J. Zhu et al., JINST 20 (2025) T01011. https://doi.org/10.1088/1748-0221/20/01/T01011

[19] Z. Geng, Nucl. Instrum. Methods Phys. Res. A 963 (2020) 163738. https://doi.org/10.1016/j.nima.2020.163738

[20] Z. Geng, R. Kalt, Nucl. Sci. Tech. 30 (2019) 10. https://doi.org/10.1007/s41365-019-0670-7

[21] R.W. Wall, "Simple methods for detecting zero crossing," in Proc. IEEE IECON, 2003, pp. 502–508. https://doi.org/10.1109/IECON.2003.1280634